\documentclass[aps,prl,twocolumn,amsmath,amssymb,nofootinbib,superscriptaddress,floatfix]{revtex4-1}

\usepackage{amsmath}
\usepackage{amssymb}
\usepackage{amsthm}
\usepackage[pdftex]{color}
\usepackage{graphicx}% Include figure files
\usepackage{dcolumn} % Align table columns on decimal point
\usepackage{bm} % bold math
\usepackage[hypertex]{hyperref}
\usepackage{longtable}
\usepackage{ulem}   % to strike things out
\newcommand{\bs}[1]{{\boldsymbol{#1}}}

\begin{document}

\title{ 
Fractional quantum Hall states at zero magnetic field
      }

\author{Titus Neupert} 
\affiliation{
Condensed Matter Theory Group, 
Paul Scherrer Institute, CH-5232 Villigen PSI,
Switzerland
            } 

\author{Luiz Santos} 
\affiliation{
Department of Physics, 
Harvard University, 
17 Oxford Street, 
Cambridge, Massachusetts 02138,
USA
            } 

\author{Claudio Chamon} 
\affiliation{
Physics Department, 
Boston University, 
Boston, Massachusetts 02215, USA
            } 
            
\author{Christopher Mudry} 
\affiliation{
Condensed Matter Theory Group, 
Paul Scherrer Institute, CH-5232 Villigen PSI,
Switzerland
            } 

\date{\today}

\begin{abstract}
  We present a simple prescription to flatten isolated Bloch bands with
  a nonzero Chern number. We first show that approximate flattening of
  bands with a nonzero Chern number is possible by tuning ratios of
  nearest-neighbor and next-nearest-neighbor hoppings in the Haldane
  model and, similarly, in the chiral-$\pi$-flux square lattice model. 
  Then we show that perfect flattening can be attained with further range
  hoppings that decrease exponentially with distance. Finally, we
  add interactions to the model and
  present exact diagonalization results for a small system at 1/3 filling
  that support 
  (i) the existence of a spectral gap,
  (ii) that the ground state is a topological state, and 
  (iii) that the Hall conductance is quantized.
  
\end{abstract}

\maketitle

In a seminal paper, Haldane%
~\cite{Haldane88}  has shown that noninteracting 
electrons hopping on a honeycomb lattice
can exhibit the integer quantum Hall 
effect (IQHE) without the Landau levels induced by a uniform magnetic field, 
provided the system breaks time-reversal symmetry (TRS).
In that model, electrons hop with a real-valued
uniform nearest-neighbor (NN) amplitude of magnitude
$t^{\ }_{1}$  that preserves TRS, as well as 
complex-valued next-nearest-neighbor (NNN) 
amplitudes with the uniform magnitude $t^{\ }_{2}$
that break TRS. A non-vanishing $t^{\ }_{2}$
generically opens a band gap at the Fermi-Dirac points of graphene 
(half-filling). This band gap results in the Chern numbers 
taking opposite values of magnitude~1
on the upper and lower bands.
Consequently, the model exhibits an IQHE at half-filling. 

Given the fact that a band insulator can support the IQHE without
a magnetic field, a natural question that we address in this Letter
is whether a fractional quantum Hall effect (FQHE) is also possible 
in an interacting lattice model without a magnetic field. 
For the usual FQHE in an uniform magnetic field, 
all Landau levels share the same Chern number,
$\pm1$ depending on the orientation of the uniform magnetic field. 
Moreover, in the absence of disorder,
all Landau levels are flat (i.e., dispersionless) and thus
can accommodate, when partially filled, 
an exponentially large number of Slater determinants, from which
incompressible liquids are selected by interactions at some
special filling fractions.
Haldane's model fulfills the first ingredient for the FQHE:
nonvanishing Chern numbers for the single-particle Bloch bands. 
We are going to construct two-dimensional lattice models 
without magnetic fields that also
satisfy the second ingredient for the FQHE: band flattening.

There is a long history of flatband models. 
They have been studied since the 1970s in amorphous
semiconductors~\cite{Weaire71a,Weaire71b,Thorpe71}, and understood using
projection operators~\cite{Straley72}. More recently
they have been studied on kagome, honeycomb, and square lattices%
~\cite{Nagaosa00,Xiao03,Wu07,Bergman08,Green10,Kapit10}. 
In Ref.~\cite{Green10} flatbands were
isolated by gaps, and the question of whether it is possible to have
a flatband with nonzero Chern number was raised.
We shall answer this question affirmatively.
We then add interactions and show evidence that the many-body state is a
topological state with fractional Hall conductance at 1/3 filling.

Our starting point is two-dimensional local lattice models
describing the hopping of spinless fermions.
In the spirit of Haldane's model, 
we restrict the lattice models to those with only two Bloch bands
and enforce locality by only allowing NN and NNN hoppings.
We will show that, by varying the ratio of the NNN to NN
hoppings, we can deform the bands to make them flatter.
The characteristic measure for the flatness of a Bloch band is here
the ratio of the bandwith to the band gap.
We then show that this criterion for flatness can be saturated to 
the ideal limit of zero for the valence band by including
arbitrary range hoppings. However, the flattened single-particle 
Hamiltonian still preserves locality in the sense that
the hopping amplitudes decrease exponentially with
the distance between any two lattice sites.

Consider the noninteracting two-band Bloch Hamiltonian of the generic
form
\begin{subequations}
\label{eq:def-two-bands-H0}
\begin{equation}
H^{\ }_0:=
\sum_{\bs{k}\in\text{BZ}}
\psi^{\dagger}_{\bs{k}}
\mathcal{H}^{\ }_{\bs{k}}
\psi^{\ }_{\bs{k}}
,
\qquad
\mathcal{H}^{\ }_{\bs{k}}
:=
B^{\ }_{0,\bs{k}}\sigma^{\ }_0
+
\bs{B}^{\ }_{\bs{k}}\cdot\bs{\sigma}.
\label{eq:def-two-bands-noninteracting-H}
\end{equation}
Here, BZ stands for the Brillouin zone, 
$
\psi^{\dag}_{\bs{k}}=
\left(c^{\dag}_{\bs{k},\text{A}},c^{\dag}_{\bs{k},\text{B}}\right)
$,
where $c^{\dag}_{\bs{k},\text{s}}$ 
creates a Bloch states on sublattice $\text{s}=\text{A},\text{B}$,
and the $2\times2$ matrices
$\sigma^{\ }_0$ and $\bs{\sigma}$ are the identity matrix and the
three Pauli matrices acting on the sublattice indices. If we define
\begin{equation}
\widehat{\bs{B}}^{\ }_{\bs{k}}:=
\frac{\bs{B}^{\ }_{\bs{k}}}{|\bs{B}^{\ }_{\bs{k}}|},
\quad
\tan \phi_{\bs{k}} :=
\frac{\widehat{B}^{\ }_{2,\bs{k}}}{\widehat{B}^{\ }_{1,\bs{k}}},
\quad
\cos \theta_{\bs{k}} := \widehat{B}^{\ }_{3,\bs{k}},
\end{equation}
we can write the eigenvalues 
of Hamiltonian $\mathcal{H}^{\ }_{\bs{k}}$
as 
$\varepsilon^{\ }_{\pm,\bs{k}}=
B^{\ }_{0,\bs{k}}\pm|\bs{B}^{\ }_{\bs{k}}|$
and for the corresponding orthonormal eigenvectors
\begin{equation}
\begin{split}
&
\chi^{\ }_{+,\bs{k}}=
\begin{pmatrix}
e^{
-\mathrm{i}      \phi^{\ }_{\bs{k}}/2
  }
\cos\frac{\theta^{\ }_{\bs{k}}}{2}
\\
e^{
+\mathrm{i}\phi^{\ }_{\bs{k}}/2
  }
\sin\frac{\theta^{\ }_{\bs{k}}}{2}
\end{pmatrix},
\quad
\chi^{\ }_{-,\bs{k}}=
\begin{pmatrix}
e^{
-\mathrm{i}      \phi^{\ }_{\bs{k}}/2
  }
\sin\frac{\theta^{\ }_{\bs{k}}}{2}
\\
-e^{
+\mathrm{i}\phi^{\ }_{\bs{k}}/2
   }
\cos\frac{\theta^{\ }_{\bs{k}}}{2}
\end{pmatrix}.
\end{split}
\label{eq: spectral decomposition H0 b}
\end{equation}
\end{subequations}
Two examples of Hamiltonians of the form%
~(\ref{eq:def-two-bands-noninteracting-H}) 
are the following.

\begin{figure}
\includegraphics[angle=0,scale=0.43]{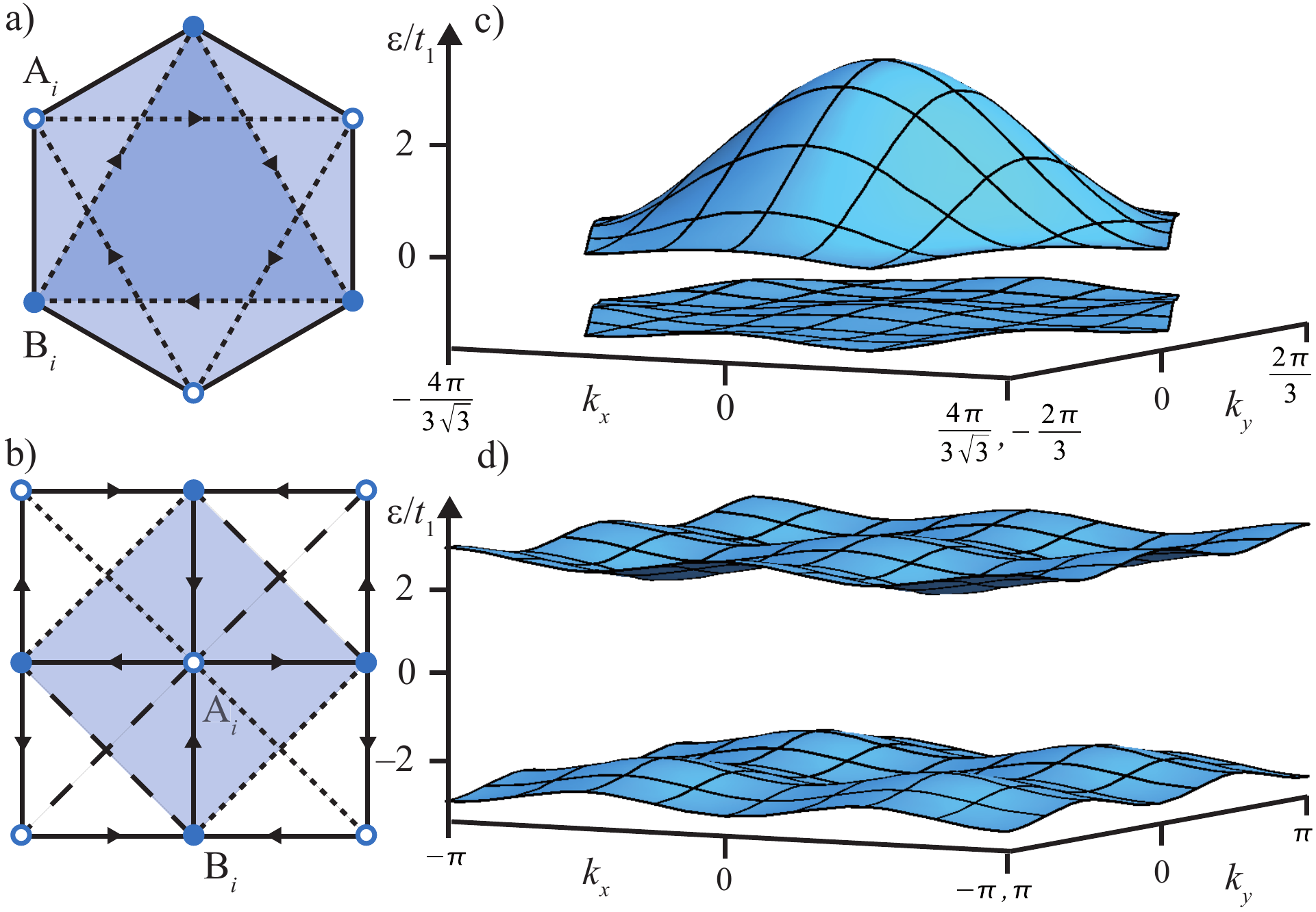}
\caption{
(Color online)
(a)
Unit cell of Haldane's model on the honeycomb lattice:
the NN hopping amplitudes 
$t^{\ }_1$ 
are real (solid lines) and
the NNN hopping amplitudes are 
$t^{\ }_2e^{\text{i}2\pi\Phi/\Phi^{\ }_0}$
in the direction of the arrow (dotted lines).
The flux $3\Phi$ and $-\Phi$ penetrate the dark shaded region
and each of the light shaded regions, respectively.
For $\Phi=\pi/3$, 
the model is gauge equivalent to 
having one flux quantum per unit cell.
(b)
The chiral-$\pi$-flux on the square lattice,
where the unit cell corresponds to the shaded area.
The NN hopping amplitudes are 
$t^{\ }_1e^{\text{i}\pi/4}$
in the direction of the arrow (solid lines) and
the NNN hopping amplitudes are $t^{\ }_2$ and $-t^{\ }_2$
along the dashed and dotted lines, respectively.
(c) The band structure of Haldane's model
for $\cos\Phi=t^{\ }_1/(4t^{\ }_{2})=3\sqrt{3/43}$
with the flatness ratio 1/7.
(d) The band structure of the chiral-$\pi$-flux for
$t^{\ }_1/t^{\ }_2=\sqrt{2}$ with the flatness ratio 1/5.
The lower bands can be made exactly flat by adding 
longer range hoppings.
        }
\label{Fig: model bands}
\end{figure}

\textit{Example 1: The honeycomb lattice.---}
We introduce the vectors 
$\bs{a}^{t}_{1}=\left(0,-1\right)$,
$\bs{a}^{t}_{2}=\left(\sqrt{3}/2,1/2\right)$,
$\bs{a}^{t}_{3}=\left(-\sqrt{3}/2,1/2\right)$
connecting NN and the vectors 
$\bs{b}^{t}_{1}=\bs{a}^{t}_{2}-\bs{a}^{t}_{3}$,
$\bs{b}^{t}_{2}=\bs{a}^{t}_{3}-\bs{a}^{t}_{1}$,
$\bs{b}^{t}_{3}=\bs{a}^{t}_{1}-\bs{a}^{t}_{2}$
connecting NNN from the honeycomb lattice
depicted in Fig.%
~\ref{Fig: model bands}(a). 
We denote with $\bs{k}$ a wave vector
from the BZ of the reciprocal lattice dual to
the triangular lattice
spanned by $\bs{b}^{\ }_{1}$ and $\bs{b}^{\ }_{2}$, say.
The model is then defined by the Bloch Hamiltonian%
~\cite{Haldane88}
\begin{subequations} 
\label{eq: def Haldane's model}
\begin{eqnarray}
&&
B^{\ }_{0,\bs{k}}:= 
2 t^{\ }_{2} \cos \Phi 
\sum_{i=1}^3
\cos \bs{k}\cdot\bs{b^{\ }_i},
\\
&&
\bs{B}^{\ }_{\bs{k}}:=
\sum_{i=1}^3
\begin{pmatrix}
t^{\ }_{1} 
\cos\bs{k}\cdot\bs{a^{\ }_i}
\\
t^{\ }_{1} 
\sin\bs{k}\cdot\bs{a^{\ }_i}
\\
-2 t^{\ }_{2}
\sin \Phi \sin\bs{k}\cdot\bs{b^{\ }_i}
\end{pmatrix},
\end{eqnarray}
\end{subequations}
where $t^{\ }_{1}\geq0$ and $t^{\ }_{2}\geq0$ 
are NN and NNN hopping amplitudes, respectively, 
and the real numbers $\pm\Phi$ are the magnetic
fluxes penetrating the two halves of the hexagonal unit cell.
For $t^{\ }_{1}\gg t^{\ }_{2}$, the gap 
$\Delta\equiv
\mathrm{min}^{\ }_{\vphantom{\pm,}\bs{k}}\varepsilon^{\ }_{+,\bs{k}}
-
\mathrm{max}^{\ }_{\vphantom{\pm,}\bs{k}}\varepsilon^{\ }_{-,\bs{k}}$ 
is proportional to $t^{\ }_{2}$.
The width of the lower band is
$\delta^{\ }_{-}\equiv
\text{max}^{\ }_{\vphantom{-,}\bs{k}}\varepsilon^{\ }_{-,\bs{k}}
-
\text{min}^{\ }_{\vphantom{-,}\bs{k}}\varepsilon^{\ }_{-,\bs{k}}$.
The flatness ratio $\delta^{\ }_{-}/\Delta$ 
is extremal for the choice $\cos\Phi=t_{1}/(4t^{\ }_{2})=3\sqrt{3/43}$,
yielding an almost flat lower band with 
$\delta^{\ }_{-}/\Delta=1/7$
[see Fig.~\ref{Fig: model bands}(c)].

\textit{Example 2: The square lattice.---}
We introduce the vectors 
$\bs{x}^{t}\equiv
\left(1/\sqrt{2},1/\sqrt{2}\right)$
and
$
\bs{y}^{t}\equiv
\left(-1/\sqrt{2},1/\sqrt{2}\right)
$
connecting NNN from the square lattice as
depicted in Fig.%
~\ref{Fig: model bands}(b). 
We denote with 
$\bs{k}^{t}=\left(k^{\ }_{x},k^{\ }_{y}\right)$ 
a wave vector from the
BZ of the reciprocal lattice dual to the 
square lattice spanned by $\bs{x}$ and $\bs{y}$.
The model is then defined by the Bloch Hamiltonian%
~\cite{Wen89}
\begin{subequations} 
\label{eq: def chiral pi-flux model}
\begin{eqnarray}
&&
B^{\ }_{0,\bs{k}}:= 
0,
\\
&&
B^{\ }_{1,\bs{k}}
+
\text{i}
B^{\ }_{2,\bs{k}}
:=
t^{\ }_{1}\,e^{-\text{i}\pi/4}
\left[
1
+
e^{
+\text{i}\left(k^{\ }_{y}-k^{\ }_{x}\right)
  }
\right]
\\
&&
\hphantom{B^{\ }_{1,\bs{k}}+\text{i}B^{\ }_{2,\bs{k}}:=}
+
t^{\ }_{1}\,e^{+\text{i}\pi/4}
\left[
e^{
-\text{i}k^{\ }_{x}
  }
+
e^{
+\text{i}k^{\ }_{y}
  }
\right],
\nonumber\\
&&
B^{\ }_{3,\bs{k}}
:=
2t^{\ }_{2}
\left(\cos k^{\ }_{x}-\cos k^{\ }_{y}\right),
\end{eqnarray}
\end{subequations}
where $t^{\ }_{1}\geq0$ and $t^{\ }_{2}\geq0$ 
are NN and NNN hopping amplitudes, respectively.
The flatness ratio $\delta^{\ }_{-}/\Delta$ 
is extremal for the choice 
$t^{\ }_{1}/t^{\ }_{2}=\sqrt{2}$, 
yielding two almost flat bands with 
$\delta^{\ }_{-}/\Delta\approx1/5$
[see Fig.~\ref{Fig: model bands}(d)].

The Chern numbers for the 
bands labeled by $\pm$ in Eq.%
~(\ref{eq: spectral decomposition H0 b}) 
are given by
\begin{equation}
C^{\ }_\pm=
\mp\!\!\!
\int\limits_{\bs{k}\in\text{BZ}}\!\!\!
\frac{\text{d}^2\bs{k}}{4\pi}\,
\epsilon_{\mu\nu}
\left[
\partial^{\ }_{k^{\ }_{\mu}}\cos\theta(\bs{k})
\right]
\left[
\partial^{\ }_{k^{\ }_{\nu}}\phi(\bs{k})
\right].
\label{eq:Cpm}
\end{equation}
They have opposite signs if nonzero. All the information about the
topology of the Bloch bands of a gaped system is encoded in the 
single-particle wave functions.  
For example, the Chern numbers depend solely on the
eigenfunctions. Haldane's model%
~(\ref{eq: def Haldane's model})
and the chiral-$\pi$-flux%
~(\ref{eq: def chiral pi-flux model}) 
are topologically equivalent 
in the sense that both have two bands with Chern numbers $\pm1$.

To enhance the effect of interactions, 
highly degenerate (i.e., flat) bands are desirable. 
It is always possible to deform the Bloch Hamiltonian%
~(\ref{eq:def-two-bands-noninteracting-H})
so as to have one flatband with the energy $-1$, say, 
while preserving the eigenspinors $\chi^{\ }_{\pm,\bs{k}}$%
~(\ref{eq: spectral decomposition H0 b}). 
Indeed, this is achieved
by turning the Bloch Hamiltonian%
~(\ref{eq:def-two-bands-noninteracting-H})
into
\begin{equation}
\mathcal{H}^{\text{flat}}_{\bs{k}}:=
\frac{\mathcal{H}^{\ }_{\bs{k}}}{
\varepsilon^{\ }_{-,\bs{k}}}.
\label{eq: def flat chiral pi flux H}
\end{equation}
Note that whenever $B^{\ }_{0,\bs{k}}\equiv 0$, 
the Hamiltonian~(\ref{eq:def-two-bands-noninteracting-H})  
has the spectral symmetry 
$\varepsilon^{\ }_{+,\bs{k}}=
-\varepsilon^{\ }_{-,\bs{k}}$
so that both bands of 
$\mathcal{H}^{\text{flat}}_{\bs{k}}=
\widehat{\bs{B}}^{\ }_{\bs{k}}\cdot\bs{\sigma}$ 
are completely flat.
This spectral symmetry applies to the 
chiral-$\pi$-flux%
~(\ref{eq: def chiral pi-flux model})  
but not to Haldane's model%
~(\ref{eq: def Haldane's model})
unless $\Phi=\pm\pi/2$.

Generically, 
$\mathcal{H}^{\text{flat}}_{\bs{k}}$ 
follows from a lattice model for which the hopping amplitudes
are nonvanishing for arbitrary large separations.
If, however, the hopping amplitudes decrease sufficiently fast 
with the separation, locality is preserved.
To estimate the decay of the hopping amplitudes 
with the separation between any two sites in the flattened
chiral-$\pi$-flux model%
~(\ref{eq: def chiral pi-flux model}), 
we calculate the decay of the Fourier coefficients $A^{\ }_{n,n'}$ of 
\begin{equation}
\frac{1}{|\varepsilon^{\ }_{\pm,\bs{k}}|}=
\sum_{n,n'=0}^\infty A^{\ }_{n,n'} \cos n k^{\ }_+ \cos n' k^{\ }_-,
\end{equation}
where $k^{\ }_{\pm}\equiv k^{\ }_{x}\pm k^{\ }_{y}$.
Because
\begin{equation}
\begin{split}
\varepsilon^{2}_{\pm,\bs{k}}=&\,
\left(2 t^{2}_{1}-t^{2}_{2}\right)
\left(2+\cos k^{\ }_{+}+\cos k^{\ }_{-}\right)
\\
&+
t^{2}_{2}\left(3+\cos k^{\ }_{+}\cos k^{\ }_{-}\right),
\end{split}
\end{equation}
it is sufficient to consider the Fourier coefficients 
$\widetilde{A}^{\ }_{n}$ 
of
\begin{equation}
\frac{1}{\sqrt{1+\alpha\cos k}}=
\sum_{n=0}^{\infty} 
\widetilde{A}^{\ }_{n} 
\cos n k,
\qquad 
\text{for}\  
-1<\alpha<1.
\label{eq:C expansion}
\end{equation} 
In the limit $n\gg1$, one finds
that $\widetilde{A}^{\ }_{n}$ decays
exponentially with $n$.
We conclude that for any fixed $n$, 
the coefficients $A^{\ }_{n,n'}$ decay exponentially with $n'$ and 
vice versa for $n$, iff $|\alpha|<1$. 
Flattening the energy bands preserves the
locality of the chiral-$\pi$-flux model%
~(\ref{eq: def chiral pi-flux model}).

The fact that we have engineered single-particle wavefunctions in a flat
Bloch band that support a Chern number $\pm 1$ is one step in mimicking the 
FQHE. However, because of lattice effects, 
it is not a given that interactions lead to a many-body ground state 
supporting the FQHE.
In fact, even when there is a uniform magnetic field, 
the combination of lattice effects and interactions 
is not well understood upon increasing
the magnetic flux threading the elementary lattice unit cell.
In Refs.%
~\cite{Sorensen05,Hafezi07},
for example,
the possibility of a FQHE induced by interactions 
for the Hofstadter problem, NN hopping
with a uniform flux threading each elementary plaquette
of the square lattice, was studied numerically. 
While the overlap between
the Laughlin states on the torus
and the lattice many-body ground states
was close to unity when the plaquette flux is 
much smaller than the flux quantum,
this overlap decreases 
when the plaquette flux becomes of the order of one quarter
of the flux quantum.
It is thus imperative to 
study how interactions lift 
the macroscopic degeneracy of a fractionally filled flat Bloch band
and whether a gapped topological ground state emerges.

Two distinctive properties of such a ground state $|\Psi\rangle$ at filling fraction $\nu$ (where $\nu^{-1}$ is an odd integer) and with periodic boundary conditions 
(toroidal geometry) are
(i) the $\nu^{-1}$-fold topological degeneracy of the ground state manifold and
(ii) the quantization of the Hall conductance $\sigma^{\ }_{xy}$ in units of 
$\nu e^{2}/h$.
The Hall conductance is related to the Chern-number $C$ 
of the many-body ground state $|\Psi\rangle$ as
$\sigma^{\ }_{xy}=Ce^{2}/h$~\cite{Niu85}.
Conventionally, the Chern-number is evaluated using twisted boundary conditions
$\langle \bs{r}+N^{\ }_x \bs{x}|\Psi^{\ }_{\bs{\gamma}}\rangle=
 e^{\text{i}\gamma^{\ }_x}\langle\bs{r}|\Psi^{\ }_{\bs{\gamma}}\rangle$ 
and
$\langle \bs{r}+N^{\ }_y \bs{y}|\Psi^{\ }_{\bs{\gamma}}\rangle=
 e^{\text{i}\gamma^{\ }_y}\langle\bs{r}|\Psi^{\ }_{\bs{\gamma}}\rangle$,
where $\bs{\gamma}^{t}=(\gamma^{\ }_x,\gamma^{\ }_y)$ 
is the twisting angle and $N^{\ }_x\times N^{\ }_y$ the number of unit cells.
The Chern number is then given by~\cite{Niu84} 
\begin{subequations}
\begin{equation}
C=\frac{1}{2\pi \text{i}}
\int\limits_{\bs{\gamma}\in[0,2\pi]^2}\!\!\!
\text{d}^2\,\bs{\gamma}\,
\bm{\nabla}^{\ }_{\bs{\gamma}}\wedge
\left\langle\Psi^{\ }_{\bs{\gamma}}\left|
\bm{\nabla}^{\ }_{\bs{\gamma}}
\right|\Psi^{\ }_{\bs{\gamma}}\right\rangle.
\label{eq:C1}
\end{equation}
Here, we introduce 
\begin{equation}
\widetilde{C}=\frac{1}{2\pi \text{i}}
\int\limits_{\bs{k}\in\text{BZ}}\!\!\!
\text{d}^2\bs{k}\;
n^{\ }_{-,\bs{k}}
\left[\bm{\nabla}^{\ }_{\bs{k}}
\wedge
\left(
\chi^{\dagger}_{-,\bs{k}}\bm{\nabla}^{\ }_{\bs{k}}\chi^{\ }_{-,\bs{k}}
\right)\right]
\label{eq:C2}
\end{equation}
\end{subequations}
as a second way to calculate the Chern number,
where $n^{\ }_{-,\bs{k}}
=\langle\Psi | c^{\dag}_{\bs{k},-}c^{\ }_{\bs{k},-}|\Psi\rangle$
is the occupation number of
the single-particle Bloch state in the lower ($-$) band
with wave vector $\bs{k}$ evaluated in the many-body ground state.
We can show that both formulas are equivalent, i.e., $C=\widetilde{C}$.
To this end, one expands the many-body wave function $|\Psi\rangle$ in 
a sum over Slater determinants and applies a gauge transformation 
to the single-particle states
to remove the twist in the boundary conditions.

We close this Letter with an exact diagonalization study
to show the existence of a gapped topological ground state
for the chiral $\pi$-flux phase~\eqref{eq: def chiral pi-flux model} 
in the presence of interactions.
We consider 
an interaction defined by the repulsive
two-body NN potential 
$V^{\ }_{i,j}$ according to
\begin{equation}
H^{\ }_{\text{int}}:=
\frac{1}{2}
\sum_{i,j\in\Lambda} 
\rho^{\ }_{i} V^{\ }_{i,j} \rho^{\ }_{j}\equiv
V
\sum_{\langle ij\rangle} 
\rho^{\ }_{i}\rho^{\ }_{j},\qquad
V>0
.
\label{eq:H-int}
\end{equation}
Directed NN bonds of the square lattice 
$\Lambda=\text{A}\cup\text{B}$
made of the open and filled circles of
Fig.%
~\ref{Fig: model bands}(b)
are here denoted by $\langle ij\rangle$,
while $\rho^{\ }_{i}$ is the occupation number on the 
site $i\in\Lambda$.

We also drive the model trough a topological phase transition to establish that 
a gapped topological many-body ground state
results from the topological nature of the model.
To this end, we add a sublattice-staggered chemical potential 
$4\mu^{\ }_{\text{s}}$
to the single-particle Hamiltonian%
~\eqref{eq:def-two-bands-H0} by replacing
$B^{\ }_{3,\bs{k}} \to B^{\ }_{3,\bs{k}}+4\mu^{\ }_{\text{s}}$
in Eq.%
~\eqref{eq: def chiral pi-flux model}.
Then, the two noninteracting bands have a 
Chern number $\pm1$ for 
$\left|t^{\ }_2/\mu^{\ }_{\text{s}}\right|>1$
and a vanishing Chern number for
$\left|t^{\ }_2/\mu^{\ }_{\text{s}}\right|<1$.
The topological phase transition at $\left|t^{\ }_2/\mu^{\ }_{\text{s}}\right|=1$
forces the single-particle spectral gap to close 
and the flattening of the bands is ill-defined 
at that point [$\alpha=1$ in Eq.~\eqref{eq:C expansion}].

\begin{figure}
\includegraphics[angle=0,scale=0.57]{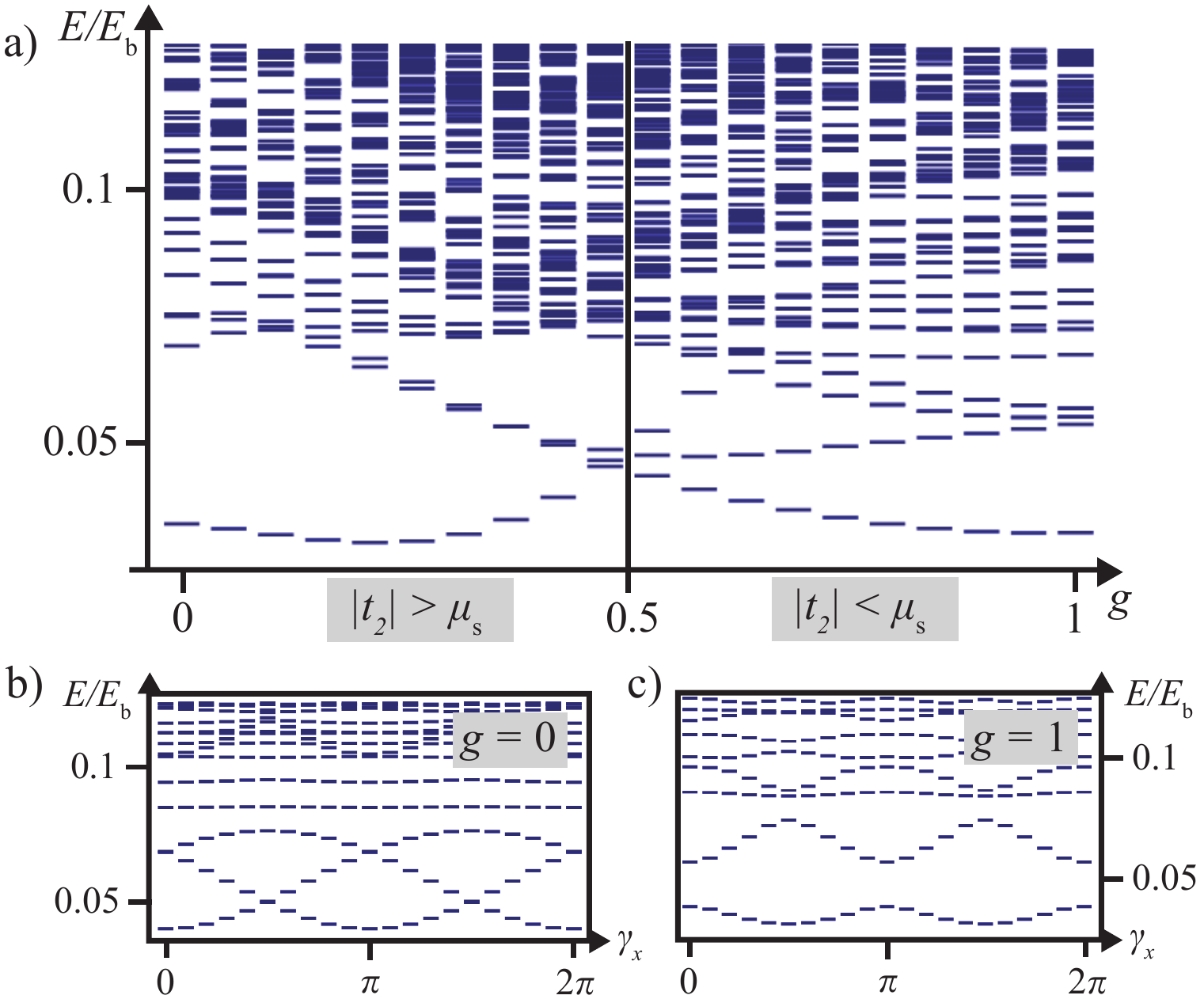}
\caption{
(Color online)
(a)
The lowest eigenvalues of $H^{\text{flat}}_0+H^{\ }_{\text{int}}$
for the chiral $\pi$-flux phase
obtained from exact diagonalization 
for 6 particles on a $3\times6$ sublattice $A$ ($1/3$ filling), 
normalized by the bandwidth $E^{\ }_{\text{b}}$.
The parameters $t^{\ }_{2}$ and $\mu^{\ }_{\text{s}}$ 
of $H^{\text{flat}}_{0}$ 
interpolate between topological ($|t^{\ }_{2}|>|\mu^{\ }_{\text{s}}|$)
and nontopological ($|t^{\ }_{2}|<|\mu^{\ }_{\text{s}}|$)
single-particle bands. 
Here, $g:=(2/\pi)\arctan|t^{\ }_{2}/\mu^{\ }_{\text{s}}|$ 
and the energies are measured relative to the energy of the
single-particle band.
(b)
The lowest eigenvalues in the center of mass momentum 
sector of the ground state of $H^{\text{flat}}_0+H^{\ }_{\text{int}}$
with twisted boundary conditions
as a function of the twisting angle $\gamma^{\ }_x$
for $\mu^{\ }_{\text{s}}=0$, $t^{\ }_{2}=t^{\ }_1/\sqrt{2}$.
The level crossings indicate the topological nontrivial nature of the
three lowest states.
(c)
Same as (b), but for $\mu^{\ }_{\text{s}}=t^{\ }_1/\sqrt{2}$, $t^{\ }_{2}=0$.
The ground state is topologically trivial.
        }
\label{Fig: energy diagrams}
\end{figure}

For a $3\times6$ sublattice $A$,
we find a unique ground state that is separated by a gap of the 
order of the interaction strength $V$ at filling $\nu=1/3$ 
[see Fig.~\ref{Fig: energy diagrams}(a)].
This state loses its clear separation in energy from the other states 
at the topological phase transition 
$\left|t^{\ }_2/\mu^{\ }_{\text{s}}\right|=1$
and another gapped ground state is obtained for 
$\left|t^{\ }_2/\mu^{\ }_{\text{s}}\right|<1$.
Based on (i) the topological degeneracy and
(ii) the quantized Hall conductance,
we will now argue that the first state is a topological many-body state 
while the latter state is topological trivial.

(i) Because of translational invariance, the Hamiltonian does not couple 
states with different center of mass momenta 
$\bs{Q}:=\bs{k}^{\ }_1+\ldots+\bs{k}^{\ }_N$, 
where $\bs{k}^{\ }_i,\ i=1,\cdots,N$ are the single-particle momenta of an 
$N$-particle state.
At 1/3 filling of the $3\times6$ sublattice $A$, the particle number $N=6$ is 
commensurate with the lattice dimensions
and all three topological states have the same $\bs{Q}$. 
As a consequence, their topological degeneracy is lifted
and a unique ground state appears. 
We can now use twisted boundary conditions to probe the topological nature 
of the ground state: 
varying $\gamma^{\ }_x$ between $0$ and $2\pi$ is equivalent to 
the adiabatic insertion of a flux quantum
in the system. During this process, a topological ground state with 
$C=1/3$ should undergo two level crossings
with the other two gapped topological states~\cite{Thouless89}.
Indeed, we find these level crossings 
for the gapped ground state when the model has topological 
single-particle bands [Fig.~\ref{Fig: energy diagrams}(b)],
whereas no level crossings are found otherwise 
[Fig.~\ref{Fig: energy diagrams}(c)].

(ii) We have also calculated the Chern number of the gapped ground state 
for $\mu^{\ }_{\text{s}}=0$, $t^{\ }_{2}=t^{\ }_1/\sqrt{2}$
with the two equivalent formulas~\eqref{eq:C1} and ~\eqref{eq:C2}. 
We find $C=0.29$ and $\widetilde{C}=0.30$
and attribute the deviations from $C=1/3$ 
to the limitations of the small system size.
For the topological trivial model with 
$\mu^{\ }_{\text{s}}=t^{\ }_1/\sqrt{2}$, $t^{\ }_{2}=0$, 
we find that $C$ and $\widetilde{C}$ vanish to a precision 
of $10^{-6}$ and $10^{-3}$, respectively.

In summary, 
we have proposed a simple recipe to deform
any noninteracting lattice model so as to obtain flatbands,
while preserving locality.
We flattened the bands of the chiral $\pi$-flux phase
and then lifted the resulting macroscopic ground state degeneracy 
with repulsive interactions.
Via exact diagonalization, we have shown that a 
FQH-like topological ground state is obtained at 1/3 filling.
This ground state, that is not well described by Laughlin-type wavefunctions,
will be further studied in future works.

We gratefully acknowledge Rudolf Morf, 
Maurizio Storni,
and Xiao-Gang Wen
for useful discussions.
This work was supported in part by DOE Grant DEFG02-06ER46316.
TN and CM 
thank the
Condensed Matter Theory Visitor's Program at Boston
University for support.

\textit{Note added.---}
Recently, we became aware of Refs.%
~\cite{Tang10,Sun10}
in which similar topological flatband models are discussed.
Subsequently, 
Ref.~\cite{Sheng11} appeared with exact diagonalization results
that are consistent with our findings.


\begin{thebibliography}{99}

\bibitem{Haldane88} 
F. D. M. Haldane,
Phys.\ Rev.\ Lett.\ \textbf{61}, 2015 (1988).

\bibitem{Weaire71a} 
D. Weaire,
Phys.\ Rev.\ Lett. \textbf{26}, 1541 (1971).

\bibitem{Weaire71b} 
D. L. Weaire and M. F. Thorpe,
Phys.\ Rev.\ B \textbf{4}, 2508 (1971).

\bibitem{Thorpe71} 
M. F. Thorpe and D. L. Weaire,
Phys.\ Rev.\ B \textbf{4}, 3518 (1971).

\bibitem{Straley72} 
J. P. Straley,
Phys.\ Rev.\ B \textbf{6}, 4086 (1972).

\bibitem{Nagaosa00} 
K. Ohgushi, S. Murakami, and N. Nagaosa,
Phys.\ Rev.\ B \textbf{62}, R6065 (2000).

\bibitem{Xiao03} 
Y. Xiao, V. Pelletier, P. M. Chaikin, and D. A. Huse,
Phys.\ Rev.\ B \textbf{67}, 104505 (2003).

\bibitem{Wu07} 
C. Wu, D. Bergman, L. Balents, and S. Das Sarma,
Phys.\ Rev.\ Lett.\ \textbf{99}, 070401 (2007).

\bibitem{Bergman08} 
D. L. Bergman, C. Wu, and L. Balents,
Phys.\ Rev.\ B \textbf{78}, 125104 (2008).

\bibitem{Green10}
D. Green, L. Santos, and C. Chamon,
Phys.\ Rev.\ B \textbf{82}, 075104 (2010).

\bibitem{Kapit10}
E. Kapit and E. Mueller,
Phys.\ Rev.\ Lett.\ \textbf{105}, 215303 (2010).

\bibitem{Wen89}
X. G. Wen, F. Wilczek, and A. Zee, 
Phys.\ Rev.\ B \textbf{39}, 11413 (1989).

\bibitem{Sorensen05}
A. S. Sorensen, E. Demler, and M. D. Lukin, 
Phys.\ Rev.\ Lett.\ \textbf{94}, 086803 (2005).

\bibitem{Hafezi07}
M. Hafezi, A. S. Sorensen, E. Demler, and M. D. Lukin,
Phys.\ Rev.\ A \textbf{76}, 023613 (2007).

\bibitem{Niu85}
Q. Niu, D. J. Thouless, and Y. S. Wu,
Phys.\ Rev.\ B \textbf{31}, 3372 (1985).

\bibitem{Niu84}
Q. Niu and D. J. Thouless,
J.\ Phys.\ A \textbf{17}, 2453 (1984).

%\bibitem{Ong05}
%N. P. Ong and W.-L. Lee,
%in \textit{``Foundations of Quantum Mechanics in the Light of New
%Technology''} (Proceedings of ISQM Tokyo 2005), 
%edited by S. Ishioka and K. Fujikawa
%(World Scientific, Singapore).

\bibitem{Thouless89}
D. J. Thouless, 
Phys.\ Rev.\ B \textbf{40}, 12034 (1989).

%\bibitem{Laughlin83}
%R.\ B.\ Laughlin, 
%Phys.\ Rev.\ Lett.\ \textbf{50}, 1395 (1983).

%\bibitem{Haldane83}
%F. D. M. Haldane, 
%Phys.\ Rev.\ Lett.\ \textbf{51}, 605 (1983).

%\bibitem{Charrier10}
%D. Charrier and C. Chamon,
%Ann.\ Phys.\ \textbf{325}, 185 (2010).

\bibitem{Tang10}
E. Tang, J. W. Mei, and X. G. Wen,
arXiv:1012.2930 (unpublished).

\bibitem{Sun10}
K. Sun, Z. Gu, H. Katsura, and S. Das Sarma,
arXiv:1012.5864 (unpublished).

\bibitem{Sheng11}
D. N. Sheng, Z. Gu, K. Sun, and L. Sheng,
arXiv:1102.2658 (unpublished).


\end{thebibliography}
\end{document}